\documentclass[12pt]{article}

\usepackage{amsmath}
\usepackage{amssymb}
\usepackage[dvips]{graphicx}
\usepackage{cite}

\makeatletter
 
  \@addtoreset{equation}{section}
 \makeatother

\newcommand {\n}{\nonumber \\}
\newcommand {\tr}{\mbox{tr}}

\begin{document}
\setlength{\oddsidemargin}{0cm}
\setlength{\baselineskip}{7mm}

\begin{titlepage}

~~\\

\vspace*{0cm}
    \begin{Large}
       \begin{center}
         {Extension of IIB Matrix Model by Three-Algebra}
       \end{center}
    \end{Large}
\vspace{1cm}

\begin{center}
           Matsuo S{\sc ato}\footnote
           {
e-mail address : msato@cc.hirosaki-u.ac.jp}\\
      \vspace{1cm}
       
         {\it Department of Natural Science, Faculty of Education, Hirosaki University\\ 
 Bunkyo-cho 1, Hirosaki, Aomori 036-8560, Japan}

\end{center}

\hspace{5cm}

\begin{abstract}
\noindent

We construct a Lie 3-algebra extended model of the IIB matrix model. It admits any Lie 3-algebra and possesses the same supersymmetry as the original matrix model, and thus as type IIB superstring theory. We examine dynamics of the model by taking minimal Lie 3-algebra that includes u(N) Lie algebra as an example. There are two phases in the minimally extended model at least classically. The extended action reduces to that of the IIB matrix model in one phase. In other phase, it reduces to a more simple action, which is rather easy to analyze.

\end{abstract}

\vfill
\end{titlepage}
\vfil\eject

\setcounter{footnote}{0}

\section{Introduction}
\setcounter{equation}{0}
The IIB matrix model \cite{IKKT} is a convincing candidate of non-perturbative type IIB superstring theory. It possesses the same space-time supersymmetry as the corresponding string theory. The existence of graviton in this model is guaranteed by the thirty-two supercharges that generate the $\mathcal{N}=2$ supersymmetry in ten dimensions.

Although the IIB matrix model is known to give some perturbative and non-perturbative dynamics in string theory, it is difficult to examine the matrix model because of its numerous interactions. We need many ideas to study non-perturbative aspects of string theory. Extending the model is one way to provide new ideas to understand not only non-perturbative aspects of string theory but also the original matrix model.

In this paper, we construct a supersymmetric model that is an extension of the IIB matrix model. "Extension" means that the model possesses a 3-algebraic structure \cite{Nambu, Filippov, BergshoeffSezginTownsend, deWHN, Yoneya, Minic, Smolin1, Smolin2, Azuma, Nogo1, Jabbari, BLG1, Gustavsson, BLG2, Lorentz0, GomisSalimPasserini, HosomichiLeeLee, Nogo2, sp1, sp2, Nogo3, Lorentz1, Lorentz2, Lorentz3, Iso, ABJM, N=6BL, sp3, PangWang, ABJ, text, Nogo8, NishinoRajpoot, kac, bosonicM, Class, GustavssonRey, HanadaMannelliMatsuo, IshikiShimasakiTsuchiya, KawaiShimasakiTsuchiya, IshikiShimasakiTsuchiya2, MModel, DeBellisSaemannSzabo, Palmkvist, LorentzianM, ZariskiM} that includes a Lie-algebraic structure\footnote{
Such an extension of the BFSS matrix theory is studied in \cite{3-algBFSS}.}. The extended model admits any Lie 3-algebra, whose triple product is totally antisymmetric. It possesses the same supersymmetry as the IIB matrix model, namely as type IIB superstring theory.  Therefore, the existence of graviton is guaranteed in the extended model. 

We also study dynamics of the extended model by taking the minimal Lie 3-algebra that includes $u(N)$ algebra as a concrete example. In this model, we find that there are two phases at least classically. While it reduces to the original IIB matrix model in one phase, it reduces to a more simple model in another phase. This simple model is rather easy to examine.

The organization of the rest of this paper is as follows.  In section 2, we construct an extended IIB matrix Model that admits any Lie 3-algebra. In section 3, we choose the minimal Lie 3-algebra that includes $u(N)$ algebra and study the model with that algebra as a concrete example. In section 4, we elucidate a phase structure of the extended model. In section 5, we conclude and discuss on the model.

\vspace{1cm}

\section{Extended IIB Matrix Model}
The IIB matrix model possesses the same supersymmetry as type IIB superstring theory, namely chiral $\mathcal{N}=2$ spacetime supersymmetry in ten dimensions \cite{IKKT}. The model also possesses u(N) gauge symmetry, which is conjectured to include diffeomorphism of the ten-dimensional spacetime. In this section, we construct an action that possesses the same supersymmetry as above. 

Let us consider the Lie 3-algebra-valued scalar $\Phi$, SO(1,9) vector $X^M$ $(M = 0, \cdots 9)$ and SO(1,9) Majorana-Weyl fermion $\Theta$ that satisfies $\Gamma^{10}\Theta=-\Theta$.

We examine dynamical supertransformations,
\begin{eqnarray}
&&\delta X^{M} = i \bar{E} \Gamma^{M} \Theta \n
&&\delta \Phi = 0 \n
&&\delta \Theta = \frac{i}{2} [\Phi, X_{M}, X_{N}] \Gamma^{MN} E,
\label{SUSY1}
\end{eqnarray}
where
\begin{equation}
\Gamma^{10}E=-E.
\end{equation}
We construct a supersymmetric model by assuming no specific Lie 3-algebra but by using only totally antisymmetry of the three products, invariance of the metric and the fundamental identity, which is an analogue of Jacobi identity.  

This transformations give a supersymmetry algebra,
\begin{eqnarray}
&&(\delta_2 \delta_1 - \delta_1 \delta_2)\Phi = 0 \n
&&(\delta_2 \delta_1 - \delta_1 \delta_2)X^{M} = 
-2 \bar{E}_2 \Gamma_{N} E_1 [\Phi, X^{N}, X^{M}] \n
&&(\delta_2 \delta_1 - \delta_1 \delta_2)\Theta = 
-2 \bar{E_2} \Gamma_{N} E_1  [\Phi, X^{N}, \Theta] \n
&& \qquad \qquad \qquad \qquad +(\frac{7}{8}\bar{E_2} \Gamma_{L} E_1 \Gamma^{L} -\frac{1}{8} \bar{E_2} \Gamma_{L_1 L_2 L_3 L_4 L_5} E_1 \Gamma^{L_1 L_2 L_3 L_4 L_5})
[\Phi, X_{N} \Gamma^{N}, \Theta]. \n
&&
\end{eqnarray}
The right-hand sides of second and third lines represent gauge transformation of $X^M$ and $\Theta$, respectively. This algebra closes on-shell if we treat 
\begin{equation}
[\Phi, X_{N} \Gamma^{N}, \Theta]=0
\end{equation}
as the equation of motion of $\Theta$. 
By transforming this equation under (\ref{SUSY1}), we have the equation of motion of $X^M$, 
\begin{equation}
[\Phi, X^{N}, [\Phi, X_{N}, X^{M}]] - \frac{1}{2} [\Phi, \bar{\Theta} \Gamma^{M}, \Theta] = 0.
\end{equation}
These equations of motion are derived from 
\begin{equation}
S=< -\frac{1}{4}[\Phi, X^{M}, X^{N}]^2 
+\frac{1}{2}\bar{\Theta} \Gamma^{M} [\Phi, X_{M}, \Theta] >.
\label{extIIBaction}
\end{equation}
This action is invariant under (\ref{SUSY1}). There is no ghost in this model because this action does not include the elements of a center. 

There is another supersymmetry of the action (\ref{extIIBaction}), which is called kinematical supersymmetry, generated by
\begin{equation}
\tilde{\delta} \Theta = \tilde{E},
\end{equation}
and the other fields are not transformed.

Let us summarize the supersymmetry algebra which ($\ref{extIIBaction}$) possesses. First, the commutator of the dynamical supersymmetry transformations gives
\begin{eqnarray}
&&(\delta_2 \delta_1 - \delta_1 \delta_2)\Phi = 0 \n
&&(\delta_2 \delta_1 - \delta_1 \delta_2)X^{M} = 0\n
&&(\delta_2 \delta_1 - \delta_1 \delta_2)\Theta =0
\end{eqnarray}
on-shell and up to the gauge symmetry. The commutator of the dynamical ones gives
\begin{eqnarray}
&&(\tilde{\delta}_2 \tilde{\delta}_1- \tilde{\delta}_1 \tilde{\delta}_2) \Phi = 0 \n
&&(\tilde{\delta}_2 \tilde{\delta}_1- \tilde{\delta}_1 \tilde{\delta}_2) X^M =0 \n
&&(\tilde{\delta}_2 \tilde{\delta}_1- \tilde{\delta}_1 \tilde{\delta}_2) \Theta =0.
\end{eqnarray}
The commutator of the dynamical and kinematical ones gives
\begin{eqnarray}
&&(\tilde{\delta}_2 \delta_1- \delta_1 \tilde{\delta}_2) \Phi = 0 \n
&&(\tilde{\delta}_2 \delta_1- \delta_1 \tilde{\delta}_2) X^M = i \bar{E}_1 \Gamma^M E_2 \n
&&(\tilde{\delta}_2 \delta_1- \delta_1 \tilde{\delta}_2) \Theta = 0.
\end{eqnarray}
If we define
\begin{eqnarray}
&&\delta' = \delta + \tilde{\delta} \n
&&\tilde{\delta}' = i(\delta - \tilde{\delta}),
\end{eqnarray}
we obtain 
\begin{eqnarray}
&&(\delta_2' \delta_1'- \delta_1' \delta_2') \Phi = 0 \n
&&(\delta_2' \delta_1'- \delta_1' \delta_2')  X^M = i \bar{E}_1 \Gamma^M E_2 \n
&&(\delta_2' \delta_1'- \delta_1' \delta_2')  \Theta = 0,
\end{eqnarray}
\begin{eqnarray}
&&(\tilde{\delta}_2' \tilde{\delta}_1'- \tilde{\delta}_1' \tilde{\delta}_2') \Phi = 0 \n
&&(\tilde{\delta}_2' \tilde{\delta}_1'- \tilde{\delta}_1' \tilde{\delta}_2')  X^M = i \bar{E}_1 \Gamma^M E_2 \n
&&(\tilde{\delta}_2' \tilde{\delta}_1'- \tilde{\delta}_1' \tilde{\delta}_2')  \Theta = 0,
\end{eqnarray}
and
\begin{eqnarray}
&&(\tilde{\delta}_2' \delta_1'- \delta_1' \tilde{\delta}_2') \Phi = 0 \n
&&(\tilde{\delta}_2' \delta_1'- \delta_1' \tilde{\delta}_2') X^M = 0 \n
&&(\tilde{\delta}_2' \delta_1'- \delta_1' \tilde{\delta}_2') \Theta = 0.
\end{eqnarray}
This result means that 32 supertransformations $\Delta=(\delta',  \tilde{\delta}')$ form the algebra of SO(1,9) $\mathcal{N}=2$ chiral supersymmetry, which is the same supersymmetry algebra of the IIB matrix model and thus of type IIB superstring theory.

\section{Minimally Extended Model}
In the previous section, we constructed an extended model that admits any Lie 3-algebra. In the following, we consider the minimal Lie 3-algebra that includes $u(N)$ algebra. Non-zero commutators are 
 \begin{eqnarray}
&&[T^0, T^i, T^j]=[T^i, T^j]=f^{ij}_{\quad k} T^k \n
&&[T^i, T^j, T^k] = f^{ijk} T^{-},
\end{eqnarray}
where $[T^i, T^j]$ is the Lie bracket. Non-zero components of the inverse of a metric $g^{AB}=<T^AT^B>$ are given by\begin{equation}
g^{-0}=-1, \quad g^{ij}=h^{ij},
\end{equation}
where $h_{ij}$ is Cartan metric of the Lie algebra.

The action ($\ref{extIIBaction}$) with this minimal Lie 3-algebra can be written in a Lie-algebra manifest form,
\begin{eqnarray}
S&=& \tr ( 
-\frac{1}{4} (\Phi_0)^2 [X^{M}, X^{N}]^2 
-\frac{1}{2} (X_0^{M})^2[\Phi, X^{N}]^2 
+ \frac{1}{2}(X_0^{M}[X_{M}, \Phi])^2 
- \Phi_0 X_0^{M} [X_{M}, X_{N}][X^{N}, \Phi] \n
&&+\frac{1}{2} \Phi_0 \bar{\Theta} \Gamma_{M}[X^{M}, \Theta]
-\frac{1}{2} X_0^{M} \bar{\Theta} \Gamma_{M}[\Phi, \Theta]
+ \frac{1}{2} \bar{\Theta} \Gamma_{M} \Theta_0 [\Phi, X^{M}] 
-\frac{1}{2} \bar{\Theta}_0 \Gamma_{M} \Phi [X^{M}, \Theta]).
\label{manifestaction}
\end{eqnarray}

Let us consider backgrounds
\begin{eqnarray}
&& \Phi_0=\bar{\Phi}_0 \n
&& X^M_0=\bar{X}^M_0 \n 
&& \Theta_0 = 0 \n
&& \Phi_i = X^M_i = \Theta_i = 0,
\label{BPSbg}
\end{eqnarray}
where $\bar{\Phi}_0$ and $\bar{X}^M_0$ take arbitrary values. These backgrounds are BPS and thus stable. The fluctuations of $\Phi_0$, $X^M_0$ and $\Theta_0$ are zero modes around these backgrounds as one can see in (\ref{manifestaction}). Therefore, we should treat each of these backgrounds as independent vacuum and fix the zero modes. We elucidate the theories around these vacua in this and next section. In this section, we consider $\Phi_0 \neq 0$ phase. 

 Next, we study gauge symmetry of the action. The 3-algebra manifest form is 
\begin{equation}
\delta X_{\alpha} = \Lambda_{\beta \gamma} f^{\beta \gamma \delta}_{\quad \alpha} X_{\delta},
\end{equation}
where $X$ represents $X^M$, $\Phi$ and $\Theta$. We can rewrite it in the Lie algebra manifest form,
\begin{eqnarray}
&& \delta X_0 =0 \n
&& \delta X_i =  \Lambda^{(1) k}_{\quad \,\,\, i} X_k +\Lambda_i^{(2)} X_0, 
\end{eqnarray}
where 
$\Lambda^{(1) k}_{\quad \,\,\, i}=2 \Lambda_{0j} f^{jk}_{\quad i}$ and 
$\Lambda_i^{(2)}=\Lambda_{jk} f^{jk}_{\quad i}$ are independent gauge parameters. $\Lambda^{(1)}$ represents the u(N) gauge transformation, whereas $\Lambda^{(2)}$ represents a shift transformation. In $\Phi_0 \neq 0$ phase, by utilising the shift transformation, one can choose $\Phi_i=0$ gauge. In this gauge, 
\begin{equation}
[\Phi, Y, Z] = \Phi_0[Y, Z],
\end{equation}
where $Y$ and $Z$ stand for any field. As a result, we have
\begin{equation}
S = \tr( -\frac{1}{4} (\Phi_0)^2[X^M, X^N]^2 
+ \frac{1}{2} \Phi_0 \bar{\Theta} \Gamma^M[X_M, \Theta]).
\end{equation}
By field redefinitions of $X^M$ and $\Theta$, we obtain 
the action of the IIB matrix model, 
\begin{equation}
S = \tr( -\frac{1}{4} [X^M, X^N]^2 
+ \frac{1}{2} \bar{\Theta} \Gamma^M[X_M, \Theta]). \label{IIBaction}
\end{equation}
That is, in $\Phi_0 \neq 0$ phase, the minimally extended model is equivalent to the IIB matrix model.  

\section{New Supersymmetric Action}
In this section, we study $\Phi_0 = 0$ phase. In this case, the vacuum can be chosen without loss of generality as 
\begin{equation}
(\Phi_0, X^9_0, X^{\mu}_0)=(0, \bar{X}^9_0, 0), 
\end{equation}
where $\mu=0, \cdots, 8.$
In this vacuum, the action reduces to 
\begin{equation}
S=\tr(-\frac{1}{2}(\bar{X}^9_0)^2 [\Phi, X^{\mu}]^2
-\frac{1}{2}\bar{X}^9_0 \bar{\Theta} \Gamma^9 [\Phi, \Theta]).
\end{equation} 
By redefining $\bar{X}^9_0 \Phi$ as $\Phi$ and renaming $\Gamma^9$ $\Gamma$, we obtain
\begin{equation}
S=\tr(-\frac{1}{2} [\Phi, X^{\mu}]^2
-\frac{1}{2} \bar{\Theta} \Gamma [\Phi, \Theta]).
\label{newaction}
\end{equation} 
This action is invariant under a dynamical supertransformation inherited from (\ref{SUSY1}), 
\begin{eqnarray}
&&\delta \Phi = 0 \n
&&\delta X^{\mu} = i \bar{E} \Gamma^{\mu} \Theta \n
&&\delta \Theta = -i [\Phi, X_{\mu}] \tilde{\Gamma}^{\mu} E,
\end{eqnarray}
where $\tilde{\Gamma}^{\mu}=\frac{1}{2} (\Gamma \Gamma^{\mu} - \Gamma^{\mu} \Gamma)$.
The supersymmetry algebra is given by 
\begin{eqnarray}
&& (\delta_2 \delta_1 - \delta_1 \delta_2) \Phi =0 \n
&& (\delta_2 \delta_1 - \delta_1 \delta_2) X^{\mu} = 
2 \bar{E}_2 \Gamma E_1 [\Phi, X^{\mu}] \n
&& (\delta_2 \delta_1 - \delta_1 \delta_2) \Theta
=2\bar{E}_2 \Gamma E_1 [\Phi, \Theta] \n
&& \qquad \qquad \qquad \qquad +(-\frac{7}{8} \bar{E}_2 \Gamma_{M} E_1 \Gamma^M \Gamma + \frac{1}{8} \bar{E}_2 \Gamma_{M_1 M_2 M_3 M_4 M_5}E_1 \Gamma^{M_1 M_2 M_3 M_4 M_5}\Gamma)[\Phi, \Theta]. \n
&&
\end{eqnarray}
This algebra closes on-shell.
This dynamical and kinematical supersymmetries form a supersymmetry generated by 32 supercharges in the same way as in the full action. This supersymmetry is consistent with that of type IIB superstring. We can analyze this phase rather easily because (\ref{newaction}) possesses a simple form.

%
%
%

\section{Conclusion and Discussion}
\setcounter{equation}{0}
In this paper, we have constructed the extended IIB matrix model. The extended model admits any Lie 3-algebra. It possesses the same supersymmetries as the original model, thus, as type IIB superstring theory.

As an initial step to elucidate dynamics of the extended model, we have examined the model with the minimal Lie 3-algebra that includes u(N) algebra. There are BPS backgrounds (\ref{BPSbg}) specified by arbitrary fields $\Phi_0$ and $X_0^M$, and the fluctuations of $\Phi_0$, $X_0^M$ and $\Theta$ are zero modes around these backgrounds. One can treat each of the BPS configurations as independent vacuum and fix the zero modes if the BPS backgrounds are stable. As a result, at least in the classical level, we have gotten two phases. In one phase, the action reduces to the original IIB matrix model, whereas in another phase, it reduces to the new action (\ref{newaction}).

Naively, one can expect that the BPS backgrounds are exactly stable because the model possesses the maximal supersymmetry. However, some BPS backgrounds can be unstable quantum mechanically in low-dimensional field theories because of infrared divergence \cite{AIKKT}. We need to examine whether the BPS backgrounds are stable or not when quantum corrections are taken into account, by calculating an effective action of $\Phi_0$ and $X_0^M$. If the backgrounds are exactly stable, the minimally extended model has dynamics that are not described by the IIB matrix model. If the backgrounds become unstable at some order of quantum corrections, the minimally extended model is equivalent to the IIB matrix model. Even in this case, we can obtain some important dynamics of the IIB matrix model rather easily up to that order, by studying (\ref{newaction}). 

In both cases, it is important to study dynamics described by (\ref{newaction}). It is rather easy because (\ref{newaction}) has a simple form. We can discuss dimensionality of the space-time which (\ref{newaction}) determines, by calculating an effective action of eigen values of $X^M$ \cite{AIKKT} or expectation values of $\tr(X^MX^N)$ \cite{NS, KKKMS}.

In this paper, we have studied dynamics of the minimally extended model, whereas we have obtained the extended model admitting any Lie 3-algebra. In general, the extended model cannot be equivalent to the IIB matrix model. Finite-dimensional indecomposable metric Lie 3-algebras with maximally isotropic center are categorized in \cite{Class}. These algebras give  non-vanishing potentials for $\Phi_{\alpha}$ and $X^M_{\alpha}$. Thus, the extended model should possess various different dynamics.

\vspace*{1cm}

\section*{Acknowledgements}
We would like to thank T. Asakawa, K. Hashimoto, N. Kamiya, H. Kunitomo, T. Matsuo, S. Moriyama, K. Murakami, J. Nishimura, S. Sasa, F. Sugino, T. Tada, S. Terashima, S. Watamura, K. Yoshida, and especially H. Kawai and A. Tsuchiya for valuable discussions. This work is supported in part by Grant-in-Aid for Young Scientists (B) No. 25800122 from JSPS.

\end{document}